# Dynamical co-existence of excitons and free carriers in perovskite probed by density-resolved fluorescent spectroscopic method

Wei Wang[1,§], Yu Li[1,§], Xiangyuan Wang[3], Yanping Lv[3], Shufeng Wang[1,2,*], Kai Wang[3], Yantao Shi[3,*], Lixin Xiao[1], Zhijian Chen[1], and Qihuang Gong[1,2,*]

[1] *State Key Laboratory for Artificial Microstructure and Mesoscopic Physics, Department of Physics, Peking University, Beijing 100871, China.*

[2] *Collaborative Innovation Center of Extreme Optics, Shanxi University, Taiyuan, Shanxi 030006, China*

[3] *State Key Laboratory of Fine Chemicals, School of Chemistry, Dalian University of Technology, Dalian, Liaoning 116024, China*

§These authors contributed equally to this work. *Correspondence and requests for materials should be addressed to S.W. (email: wangsf@pku.edu.cn), Y.S. (email: shiyantao@dlut.edu.cn), and Q.G. (qhgong@pku.edu.cn).

**Abstract**

Using transient fluorescent spectra at time-zero, we develop a density-resolved fluorescent spectroscopic method for investigating photoproducts in $CH_3NH_3PbI_3$ perovskite and related photophysics. The density dependent dynamical co-existence of excitons and free carriers over a wide density range is experimentally observed for the first time. The exciton binding energy ($E_B$) and the effective mass of electron-hole pair can be estimated based on such co-existence. No ionic polarization is found contributing to photophysical behavior. It also solves the conflict between the large experimentally measured $E_B$ and the small predicted values. The spectroscopic method also helps to detect the true free carrier density under continuous illumination without the interference of ionic conductivity. Our methods and results profoundly enrich the study and understanding of the photophysics in perovskite materials for photovoltaic applications.

Organolead halide perovskite solar cells are promising candidates for the next generation of photovoltaic devices, which recently achieved over 20% conversion efficiency [1]. It is well-known that the rich photo-induced free carriers benefit their high efficiency [2]. Excitons were also observed co-existing with free carriers at a ratio of 1:10 [3]. However, this co-existence should be more diverse, because excitons formation by free electrons and holes may have density dependency. This indicates a dynamically co-existing relationship that controls the population ratio of excitons and free carriers. The phenomenon should be significant for perovskite since its exciton binding energy ($E_B$) is close to the thermal energy at room temperature. However, among the various models describing excited species in semiconductors, [4,5] it had not been proved whether this model is the right one. This fundamental dynamical co-existence behavior must be verified prior to further photophysical investigations. No experimental study has been conducted at the moment.

The extent of exciton ionization by thermal energy depends on $E_B$. Experimental studies revealed various $E_B$ from 98 meV down to a few meV [2,4,6-12]. There are even significant smaller $E_B$ predicted as 0.7 or 2.0 meV [13,14]. Such a wide range of $E_B$ values may be due to the widely known morphological versatility of perovskite films. Discovering sample-specific $E_B$ through a proper and convenient way is necessary for understanding the variation of $E_B$ and analyzing the effect of morphology. In the $E_B$ measurement, the ionic dielectric response should be properly included. This response is found extremely large at the low frequency limit [15,16] and may effectively reduce $E_B$ down to the small predicted values [13,17]. This ionic response is also known as the reason for the hysteresis phenomenon in devices [18-21]. New methods are required for understanding the ionic contribution to exciton formation, so as to solve the conflict between the experimental and predicted $E_B$ values. It may also significantly affect the carrier density measurement, since the widely used methods are based on the electric circuits, which will involve ionic conductivity [22,23]. New methods are urgently required to uncover the true free carrier density.

All these issues for perovskite can be largely solved by our newly developed density-resolved fluorescent spectroscopic method. The results from this method experimentally proved the dynamical co-existence of excitons and free carriers in perovskite. Such co-existence directly reveals the sample-specific $E_B$ and the effective mass ($m_{eh}$) of free electron-hole pairs. The ionic contribution in exciton formation can also be properly addressed by this method. The results solved the conflict between the experimentally measured $E_B$ and predictions. This spectroscopic method also helps to successfully achieve the true free carrier density under continuous illumination without interference from ionic conductivity. Our method and findings pave new ways for an improved understanding of the fundamental photophysics of perovskite materials and devices.

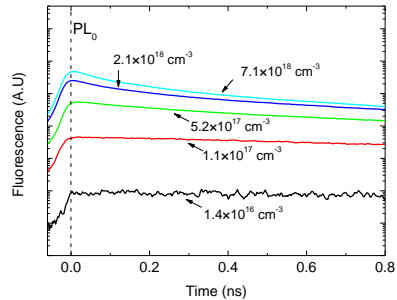

**FIG. 1.** The pump energy dependent fluorescent decay of $CH_3NH_3PbI_3$ film with a thickness of 250 nm. The pump is at 517 nm, with an injected excitation density from $1.4\times10^{16}$-$7.1\times10^{18}$ cm$^{-3}$. The $PL_0$ is taken from the maxima of the fluorescent decay with a series of excitation densities.

The CH$_3$NH$_3$PbI$_3$ perovskite films were excited by 517 nm femtosecond laser with a pulse energy between 0.13 and 68 μJ cm$^{-2}$. For films with a thickness of ~250 nm, the corresponding spatial excitation densities are from 1.4×10$^{16}$ to 7.1×10$^{18}$ cm$^{-3}$. The excitation density means the sum of excitons and free electron-hole pairs. With the increment in the pump fluence, the fluorescent decay becomes faster as shown in Fig. 1. The transient photoluminescent spectra with a maximum intensity at time zero, $PL_0$, is taken to develop the density-resolved fluorescent spectral method in our study. The weak non-zero negative signal seen in the figure is the tail end of the instrumental response, which can also be found in another report [2]. Shown in Fig. 2, the $PL_0$ intensity increases quadratically to the excitation density until 1×10$^{17}$ cm$^{-3}$. For higher pump fluence between 7×10$^{17}$ and 3×10$^{18}$ cm$^{-3}$, the $PL_0$ intensity increases linearly. Above that, the Auger process appears, where the increase of $PL_0$ is below a linear dependency. Detailed description about the method can be found in Supplemental Materials (SM).

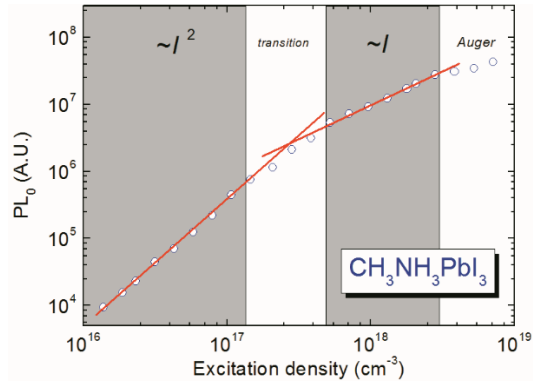

**FIG. 2.** $PL_0$ intensity *vs.* excitation density, $I(n)$. The red lines are drawn with strict quadratic and linear dependencies for comparison. The data are separated into four sections, quadratic dependency, transition, linear dependency, and the Auger process.

Among all the models, we found this dependence can be well derived by the Saha-Langmuir equation [24]. The equation described an $E_B$ controlled, density dependent balance between excitons and free carriers. For a 3D semiconductor, the equation is [25]:

$$\frac{x^2}{1-x} = \frac{1}{n}\left(\frac{2\pi\mu k_B T}{h^2}\right)^{3/2} e^{-\frac{E_B}{k_B T}} = \frac{1}{n}C(T, E_B) \tag{1}$$

in which $x$ is the ratio of free carriers among the total excitation density $n$. $\mu$ is the reduced effective mass of the exciton (~0.15 $m_e$, mass of electron [9]), $K_B$ is the Boltzmann constant, $T$ is the temperature, and $h$ is the Planck's constant.

Though this model was suggested applicable in perovskite [4], experimental verification is hard not be done with other methods. To prove it with our method, we need to modify the widely applied fluorescent intensity expression, $I(n) \propto A_1 n + A_2 n^2$, in which $I(n)$ is the fluorescent intensity, while $A_1$ and $A_2$ are the rates for mono- and bi-molecular recombination, respectively [26]. It is modified to:

$$I(n) = A(1-x)n + Bx^2 n^2 \tag{2}$$

in which $x$ is from Eq (1), and $A$ and $B$ are the coefficients of the linear and quadratic terms, respectively, including not only the decay rates, but also the signal collecting efficiency and the detector sensitivity. This modification has a clear physical meaning, indicating the separated contributions from excitons and free carriers. The contribution of the two terms depend on $x$, which is a function of $n$. At the low and high density limits, the quadratic and linear dependencies require $x \to 1$ and $x \to 0$, respectively. Then the Eq. (2) can then be written under these limits:

$$I = [A/C(T,E_B) + B]n^2 \quad (x \to 1) \tag{3}$$

$$I = [A + BC(T,E_B)]n \quad (x \to 0) \tag{4}$$

Equations (3) and (4) present quadratic and linear dependencies, which is in good agreement with our density-resolved measurement shown in Fig. 2. After fitting $PL_0$ at low and high excitation densities, we can obtain $C(T,E_B)$ by dividing the coefficient of Eq. (4) by that of Eq. (3).

$$C(T,E_B) = \frac{[A+BC(T,E_B)]}{[A/C(T,E_B)+B]} = \left(\frac{2\pi\mu k_B T}{h^2}\right)^{3/2} e^{-\frac{E_B}{k_B T}} \tag{5}$$

The calculated $C(T,E_B)$ is $2.7\pm0.2\times10^{17}$ cm$^{-3}$. In Fig. 2, this density can be regarded as the transition density from the quadratic to the linear dependency. From this $C(T,E_B)$, we calculated an $E_B$ of 24±2 meV. More results can be found in the SM.

The free carriers within this dynamical co-existence exhibit band filling behavior. With the increasing of excitation density, the filling of the conduction band forces the optical transition to shift towards a higher energy. Then the corresponding $PL_0$ fluorescent spectra blue-shift, as shown in Fig. 3(a). The bandgap shift, $\Delta E_g$, follows the Burstein-Moss band filling model, described as [27,28]:

$$\Delta E_g = \frac{\hbar^2}{2m_{eh}}(3\pi^2 n_{eh})^{\frac{2}{3}} \tag{6}$$

in which $n_{eh}$ is the free carrier density (equal to $x \times n$) and $m_{eh}$ is the reduced effective mass of a free electron-hole pair. A linear increment between $(x \times n)^{2/3}$ and the bandgap, $E_g$, can be found in Fig. 3(b). The linear fit yields $m_{eh} = 0.29 \pm 0.02\ m_e$. This $m_{eh}$ is nearly identical to the study using femtosecond transient absorption spectroscopy [27]. Though a variation in $m_{eh}$ is also found among the samples, as shown in SM, the linear dependency between $(x \times n)^{2/3}$ and $E_g$ is observed for all cases.

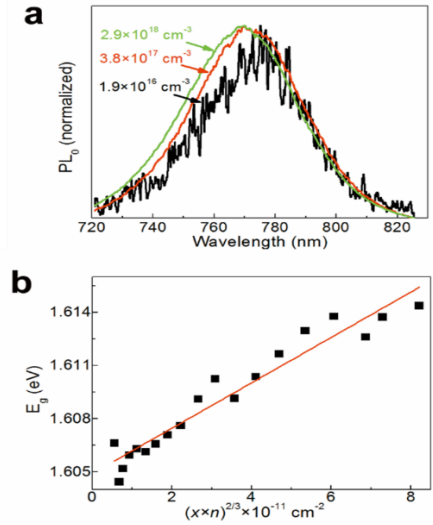

**FIG. 3.** (a) The transient emission spectra of $PL_0$ blue-shift with the increase of the pump fluence. (b) The data points are the peak positions from the Gaussian line shape fitting. The straight line is the fitting of the bandgap shift to $(x \times n)^{2/3}$, where $n$ is the total excitation density, $x$ is the ratio of free carriers among density $n$. The fitting gives a reduced effective mass of 0.29±0.02 $m_e$. This bandgap modulation is known as the Burstein-Moss band-filling effect.

In dynamical co-existence, the ratio $x$ is the function of $n$ over wide density range. A specific ratio $x$ can be figured out if we know a specific $n$ inside the material. It will be meaningful for real applications, for which the films are under continuous illumination, such as sunlight, and a static $n$ is achieved. However, for perovskite, measuring the density using an electric circuit may suffer from significant ionic conductivity. We developed a purely spectroscopic method derived from our density-resolved method to read the true free carrier density, $n_c$. The method is based on the characteristic emissive bimolecular recombination. When a film is illuminated by a continuous light, $n_c$ is stable due to the balance of excitation and recombination. Then we inject a pulsed light to introduce additional carriers with known density $n_p$. The transient fluorescent intensity, $PL_0$, is proportional to the square of $n_c+n_p$. We can estimate $n_c$ by $n_c = n_p \left(\sqrt{I_{c+p}} - \sqrt{I_p}\right)/\sqrt{I_p}$, in which $I_p$ and $I_{c+p}$ indicate $PL_0$ without and with continuous light, respectively. The $n_c$ can be verified by selecting another $n_p$ injection.

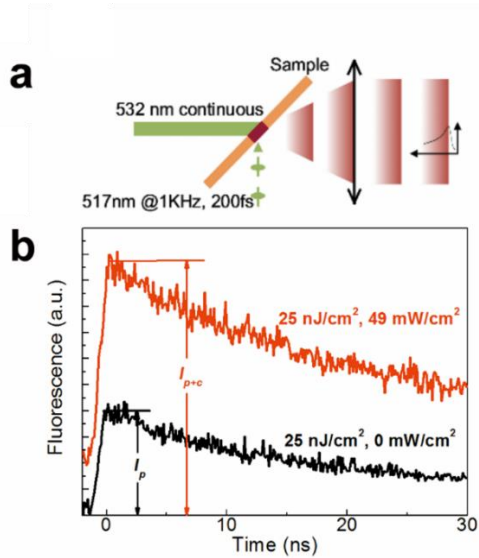

**FIG. 4.** (a) The schematic diagram of our purely optical method for carrier detection. (b) The transient fluorescent decay of pulsed carrier injection without (lower black line) and with continuous light (upper red line). The subscript $p$, and $c+p$ means pulsed light and their combination, respectively. The continuous light is 49 mWcm$^{-2}$ (~1 sun, corresponding to ~100 mW cm$^{-2}$ sunlight of UV-vis + infrared), while the pulsed light is 25 nJ cm$^{-2}$, corresponding to $n=2.6\times10^{15}$ cm$^{-3}$. The continuous and pulsed light are 532 nm and 517 nm, respectively.

Figures 4(a) and 4(b) are a schematic diagram of the method and the representative signals, respectively. The result shows a carrier density of $1.40\pm0.12\times10^{15}$ cm$^{-3}$ under a 49 mW cm$^{-2}$ continuous illumination at 532 nm (~1 sun). The correspondent carrier lifetime, $1/k$, is then estimated according to the balance between carrier injection and decay, $dn_c/dt = kn_c$, as $0.26\pm0.02$ μs. More details can be found in SM.

Our density-resolved fluorescent spectral method is designed based on the specialties of the perovskite. In perovskite, the electrons and holes are both free carriers and can co-exist at high density. It has both mono- and bimolecular emissive recombination. Ionic response in perovskite is significant comparing to other widely studied semiconductors. In addition, the photophysical behaviours strongly depend on morphology. It also shows phase-transitions at ~160K. Our method successfully make use of the specialities and overcome the difficulties encountered by other methods.

The density-resolved spectroscopic study gives a sample-specific $E_B$ of 26 meV. The value effectively include the ionic contribution. Perovskites have highly mobile ions such as methylammonium, $CH_3NH_3^+$, which was recently well addressed and observed in NMR spectra [29-31]. The ionic dielectric response can significantly screen the Coulomb interaction between the electron and hole in an exciton, and effectively reduce $E_B$. This is the physics behind the predicted extra small $E_B$. Therefore, the $E_B$ measurement should properly include the ionic contribution. Beforehand, one must keep in mind that the ionic response is non-instantaneous to the electric field according to the Franck Condon principle, because it requires nuclei motion for polarization. There are both absorption and fluorescence detection methods to study $E_B$. The widely applied band-edge absorption methods instantaneously respond to electric fields, which does not include such ionic polarization [2]. Fluorescent detections are non-instantaneous. However, they are based on the assumption that the thermal-induced exciton ionization leads to fluorescent quenching [32], while in perovskite, the ionized excitons experience emissive bimolecular recombination. In addition, they are temperature dependent methods, which may include uncertainties because of the phase transition at ~160K [12,33]. To surmount all these obstacles, new $E_B$ measurement methods should satisfy requirements such as independence from temperature, counting in both mono- and bi-molecular emission, and non-instantaneous response.

When we take $PL_0$ as the signal, it is temperature independent and includes both mono- and bi-molecular emissive recombination. It also satisfies the requirement of non-instantaneous response. In transient spectroscopy, this $PL_0$ approximately represents the integration of the excitation process (if there is no fast decay or raising component after excitation). Therefore, the pulsed excitation is located at the temporal center of the raising edge. The $PL_0$ then naturally includes a delay to the excitation, which is about half of the raising time. This delay is actually decided not by the excitation pulse width, but by the instrumental response function, such as ~30 ps in our streak camera system when the observation window is 1 ns. This is a universal characteristic for all time-resolved spectroscopy. Therefore, our $PL_0$ is non-instantaneous and is collected at a ~ 30 ps delay after excitation. This delay is sufficient to include the ionic polarization response. According to an ultrafast spectroscopic study, the formation of an exciton takes ~1 ps [34]. The phonon in perovskite system was reported to have a maximum energy of ~25 meV [35], corresponding to an oscillation period less than 200 fs. A recent study showed a ~3 ps cation reorientation motion in the iodide lattice [36]. Therefore, the stabilization of the ionic response, if it exists, is shorter than the delay of $PL_0$ by an order of magnitude. Thus, the new method properly includes the possible ionic contribution.

As the result, our $E_B$ is more than one order of magnitude larger than the predicted values. This result clearly indicates a negligible contribution from the ionic response. The prediction of an extremely small $E_B$ based on a large dielectric constant cannot be proved. This may be because that $E_B$ is not smaller than the phonon energy, or, because the cation reorientation is not fast enough to introduce an additional dielectric response. Only an instantaneous response component in $\varepsilon$ should be applied in this system. This analysis gives a clear answer to support experiments and solves the conflict with small predicted values. This also means that band-edge absorption methods are safe for estimating $E_B$ in this system.

Our carrier density detection method gives us the possibility to detect the true free carrier density without the interference of ionic conductivity. We obtained an accumulated spatial charge density of $1.40\pm0.12\times10^{15}$ cm$^{-3}$ under continuous illumination, which is an order of magnitude higher than the value measured through the Hall Effect [19]. This method provides a new way to analyze the photophysics inside perovskite through the dimension of carrier density. By knowing dynamical co-existence, we can easily learn that most of the photoproducts are free carriers under this density. We can also calculate the corresponding carrier lifetime which is found to be comparable to the fluorescence decay under low pump fluence.[37] The long lifetime and slow bimolecular recombination have been reported due to the requirement of an activation energy of 75 meV [8], and is responsible for the high charge density, which should be the reason for the high cell efficiency [38,39].

In this research, we developed a density-resolved fluorescent spectroscopic method to study the photoproducts inside perovskite and the related photophysics. This method is successful in proving the dynamical co-existence of excitons and free carriers in CH$_3$NH$_3$PbI$_3$ perovskite films. The results show that the dynamical co-existence is in good agreement with the Saha-Langmuir equation, and the Burstein-Moss band filling effect. It gives a new underlying picture of photoproducts in perovskite over a wide density range. The method provides a series of ways to study $E_B$, effective mass, ionic contribution in exciton formation, and true carrier densities without ionic conductivity, indicating the versatility of this technique. Together with the $E_B$ value and the physical meanings behind it, we proved the limited ionic contribution to exciton formation. We give a clear answer that why the predicted small $E_B$ based on a large dielectric constant is not applicable in this perovskite system. The detection of the true free carrier density under continuous illumination provides a new dimension for analyzing photophysics inside perovskite. Both the method and the physics present plentiful new information for understanding the photophysics inside perovskite. More photophysical studies can be expected to be carried out based on our density-resolved spectroscopic methods and our discoveries in physics.

This work supported by the National Basic Research Program of China 2013CB921904; National Natural Science Foundation of China under grant Nos. 11134001, 61575005，11574009, and 51402036.

Supplemental Materials

# Dynamical co-existence of excitons and free carriers in perovskite probed by density-resolved fluorescent spectroscopic method

Wei Wang[1,§], Yu Li[1,§], Shufeng Wang[1,2,*], Xiangyuan Wang[3], Kai Wang[3], Yantao Shi[3,*], Lixin Xiao[1], Zhijian Chen[1], and Qihuang Gong[1,2,*]

[1] *State Key Laboratory for Artificial Microstructure and Mesoscopic Physics, Department of Physics, Peking University, Beijing 100871, China.*

[2] *Collaborative Innovation Center of Extreme Optics, Shanxi University, Taiyuan, Shanxi 030006, China*

[3] *State Key Laboratory of Fine Chemicals, School of Chemistry, Dalian University of Technology, Dalian, Liaoning 116024, China*

§These authors contributed equally to this work. *Correspondence and requests for materials should be addressed to S.W. (email: wangsf@pku.edu.cn), Y.S. (email: shiyantao@dlut.edu.cn ), and Q.G. (qhgong@pku.edu.cn).

**Method and sample preperation**

**Spectroscopic method**

The light source is from a mode-lock Ti:sapphire femtosecond laser (Legend, Coherent) pumped two-stage optical parametric amplifier (OperA Solo, Coherent). It generates pump pulse of 517 nm at a repetition rate of 1 KHz, with pulse width of ~200 fs and energy up to 100 μJ per pulse. The temporal-spectral fluorescence were recorded with a streak camera system (Hamamatsu) centered at 780nm. Short time windows of 1 ns were applied in order to precisely resolve $PL_0$. The continuous light is 532 nm from a diode laser.

The parameters of the method need to be carefully selected. The most important thing is to select a detection system with proper response time. This response time should be significantly longer than the internal conversion process, which is at the scale of picosecond. Then the time-zero fluorescent spectra will represent the cooled photoproducts only. On the other hand, the response time should be significantly shorter than the excited state decay. This means the impact of decay is ignorable during the system response. However, the amendment of data may be necessary when the decay can not be ignored. In FIG. SM1, we show how to correct signal when the decay is counted in.

As shown in FIG. SM1(a), we present several simulated exponential fluorescent decays under ultrafast excitation. When the detection system has a slow response time, such as ~30ps, the fluorescence convolute with the instrumental response function. The results are shown in FIG. SM1(b). It is clearly shown that when the decay time is not infinite long, the time-zero fluorescent intensity will become weaker. When the fluorescent decay is 0.15 ns, 0.3 ns, and 1 ns, their fluorescent maxima are of the ratio 83%, 90%, and 96% to the one with infinite decay time, respectively. The data correction here is to convert the intensity maxima with finite decay time to the intensity with infinite decay time. This can easily be done by dividing the maxima with corresponding ratio. For real data processing, we can firstly deconvolute the real signal to find out the decay rate. Then we run a numerical simulation to figure out the ratio. It should be kept in mind that though shorter decay can also be used in such system, it is not recommended since larger uncertainty may occur in data processing. Here we suggest to keeps the decay ~5 times longer than the system response time.

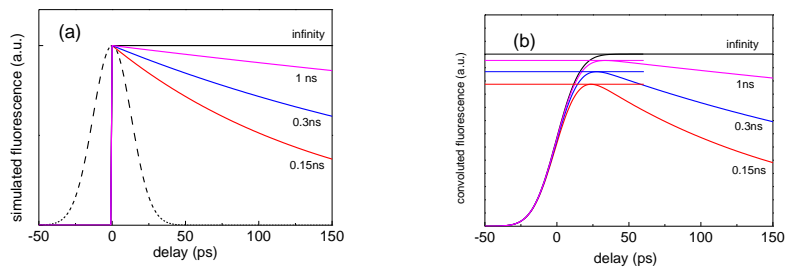

FIG SM1.  (a)The simulated fluorescent decay (0.15ns, 0.3ns, 1ns, and infinity, solid color lines) and system response function with 30ps width (segment line). (b) The convolution of similated fluorescent decay with system response function. Their maxima are marked with horizontal lines.

**CH$_3$NH$_3$PbI$_3$ films.** They were deposited on cleaned glass slides with a two-step sequential method in a nitrogen atmosphere. First, 70 μL of a hot (70 °C) solution of PbI$_2$ in DMF (N,N-Dimethylformamide, 1 M) was spin-coated on the substrate at 5000 r.p.m. for 10 s and sequentially annealed at 70 °C. After cooling to room temperature, the samples were dipped in 2-propanol for 10 s before being transferred and dipped into a CH$_3$NH$_3$I 2-propanol (10 mg/ml) solution for 2 min. Then the films are rinsed with 2-propanol and dried by spin-coating at 3000 r.p.m. for 30 s. Finally annealing was performed at 70 °C to obtain the perovskite thin films.

**The derivation of Saha-Langmuir equation for dynamical co-existence of excitons and free carriers.**

In manuscript, the most basic two equations are Saha-Langmuir equation (1) and fluorescent decay with dynamical conversion of exciton and free carrier (2):

$$\frac{x^2}{1-x} = \frac{1}{n}\left(\frac{2\pi\mu k_B T}{h^2}\right)^{3/2} e^{-\frac{E_B}{k_B T}} = \frac{1}{n}C(T, E_B) \tag{1}$$

$$I(n) = A(1-x)n + Bx^2 n^2 \tag{2}$$

From Eq. (1), when $x \to 1$, $\frac{x^2}{1-x} \to \frac{1}{1-x}$, so that Eq. (1) $\to \frac{1}{1-x} = \frac{1}{n}C(T, E_B)$. Therefore, we know $1-x = n/C(T,E_B)$. We bring this into first term of Eq (2), and $x \to 1$ into second term of Eq (2). We will obtain the equation $I = [A/C(T, E_B) + B]n^2$. This is equation (3) in manuscript.

From Eq. (1), when $x \to 0$, $\frac{x^2}{1-x} \to x^2$, so that Eq. (1) $\to x^2 = \frac{1}{n}C(T, E_B)$. We bring this into second term of Eq. (2), and $x \to 0$ into first term of Eq. (2). We will obtain the equation $I = [A + BC(T, E_B)]n$. This is equation (4) in manuscript.

The dividing of co-efficient of (4) to (3), we obtain Eq. (5) in the manuscript, which is listed below.

$$C(T, E_B) = \frac{[A + BC(T, E_B)]}{[A/C(T, E_B) + B]} = \left(\frac{2\pi\mu k_B T}{h^2}\right)^{3/2} e^{-\frac{E_B}{k_B T}} \tag{5}$$

From this Eq. (5), we can directly calculate $E_B$.

**Simulation on dynamical co-existence based on Saha-Langmuir equation**

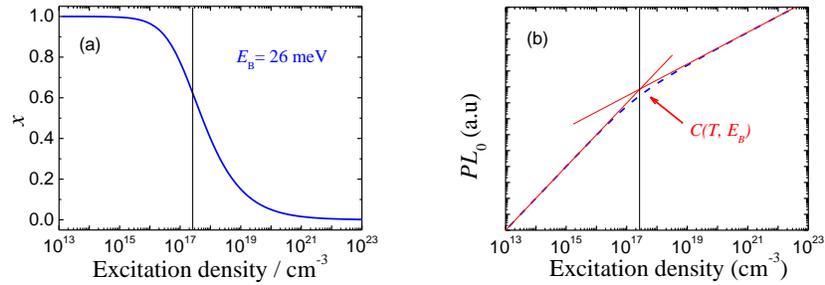

FIG. SM2. Calculated free carrier ratio $x$ and fluorescent intensity towards excitation density. (a) The ratio $x$ towards excitation density, calculated with $E_B$=26meV. The curve shows nearly unit value for $x$ at low pump fluence. The vertical line indicates the $C(T, E_B)$. (b) Simulated $PL_0$ fluorescent intensity towards excitation density (blue dash line). The red lines are drawn with strict square and linear increment for guiding. They are across at $C(T, E_B)$.

**Experiments on new samples**

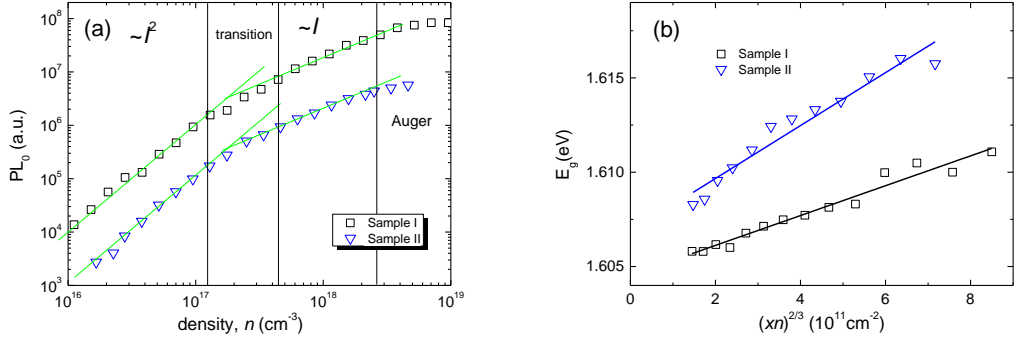

FIG. SM3. Two other samples in this investigations. (a) The quadratic to linear dependence for samples made at different batches. Highly repeatable profile can be found though the absolute intensities are different. The green lines are also drawn with strict quadratic and linear slop for reference. Small variation on $C(T, E_B)$ and binding energy are also shown, indicating sample specified variation due to inhomoginity. (b) The data points are the peak position from Gaussian line shape fitting. The straight line is the fitting of the bandgap shift to $(x \times n)^{2/3}$. $n$ is the total excitation density, while $x$ is the ratio of free carriers in $n$. The fittings give reduced effective mass. This bandgap modulation is known as Burstein-Moss band-filling effect. The result of $C(T, E_B)$, $E_B$, and effective mass ($m_{eh}$) are listed in the following table.

Table SM1. The fitting results Sample I and II in FIG SM3.

|  | Sample I | Sample II |
| --- | --- | --- |
| $C(T, E_B)$, $\times 10^{17}$ cm$^{-3}$ | 2.1 | 2.4 |
| $E_B$, meV | 32 | 29 |
| $m_{eh}$ (m$_e$, electron mass) | 0.46 | 0.26 |

Table SM2. Accumulated spatial carrier density detected by pulsed light carrier detection method.

| $I_{c,pump}$ (mW/cm$^2$) | Carrier injection (10$^{21}$·cm$^{-3}$·s$^{-1}$) | $I_{p,pump}$ (nJ/cm$^2$) | $n_p$ (10$^{15}$·cm$^{-3}$) | $F_p$ | $F_{p+c}$ | $n_c$ (10$^{15}$·cm$^{-3}$) | $\bar{n_c}$ (10$^{15}$·cm$^{-3}$) | $\overline{1/k}$ (μs) |
|---|---|---|---|---|---|---|---|---|
| 0 | 0 | 14 | 1.5 | 1.1±0.1 | | | 0 | |
| | | 25 | 2.6 | 4.0±0.3 | | | | |
| 18 | 1.9 | 14 | 1.5 | | 2.2±0.2 | 0.62±0.13 | 0.60±0.10 | 0.32±0.05 |
| | | 25 | 2.6 | | 6.0±0.4 | 0.58±0.15 | | |
| 49 | 5.3 | 14 | 1.5 | | 4.0±0.3 | 1.36±0.16 | 1.40±0.12 | 0.26±0.02 |
| | | 25 | 2.6 | | 9.6±0.4 | 1.43±0.17 | | |

$I_{c,pump}$ and Carrier injection are the intensity of continuous light at 532nm and corresponding excitation density at unit time. $I_{p,pump}$ and $n_p$ are the pulsed light energy and corresponding excitation injection per pulse. $F_p$ is the fluorescent intensity with pulsed light only. $F_{p+c}$ is the $PL_0$ when combine continuous and pulsed light. $n_c$ is the calculated spatial charge density under continuous illumination, with expression $n_c = n_p \left(\sqrt{I_{c+p}} - \sqrt{I_p}\right)/\sqrt{I_p}$. $\bar{n}_c$ is the average value of $n_c$, while $\overline{1/k}$ is the carrier lifetime calculated by $dn_c/dt = kn_c$.